\documentclass{elsart}

\usepackage{graphicx}
\usepackage{amsmath}
\usepackage{bm}         % bold math

\newcommand{\bra}[1]{\langle #1|}
\newcommand{\ket}[1]{|#1\rangle}

\begin{document}

\begin{frontmatter}
\bibliographystyle{elsart-num}

\title{Generic tripartite Bell nonlocality sudden death under local phase noise}
\author[Ann]{Kevin Ann} and
\ead{kevinann@bu.edu, correspondence author}
\author[Jaeger]{Gregg Jaeger}
\ead{jaeger@bu.edu}

\address[Ann]{Department of Physics, Boston University \\
590 Commonwealth Avenue, Boston, MA 02215}
\address[Jaeger]{Quantum Imaging Lab, Department of Electrical and Computer Engineering,\\
and Division of Natural Sciences, Boston University, Boston, MA
02215}

\date{\today}

\begin{abstract}
We definitively show, using an explicit and broadly applicable model, that 
local phase noise that is capable of eliminating state coherence only in 
the infinite-time limit is capable of eliminating nonlocality in finite 
time in three two-level systems prepared in the Bell-nonlocal tripartite 
states of the generic entanglement class. 
\end{abstract}
\begin{keyword}
% Up to six keywords permitted
Bell-type inequality, Bell nonlocality, Entanglement sudden death, multi-local dephasing 
\PACS 03.65.Ta, 03.65.Ud, 03.67.-a
% [PACS 2008 	- http://www.aip.org/pacs/pacs08/ASCII2008FullPacs.txt]
% (03.65.Ta 	- Foundations of quantum mechanics; measurement theory)
% (03.65.Ud 	- Entanglement and quantum nonlocality)
% (03.67.-a 	- Quantum information)
\end{keyword}
\end{frontmatter}

\section{Introduction}
It has recently been demonstrated that when external noise 
acts on bipartite states of compound quantum systems, a sudden 
total loss of entanglement can occur in finite time in a context where
there is persistence of some quantum coherence for {\it all} finite times, 
an effect known as Entanglement Sudden Death (ESD) 
\cite{Diosi03,DH04,YE04,YE06a,YE06b,YE07,AJ07a,JA08}.  An
explicit local hidden-variables model for entangled mixed states 
of three two-level systems has also recently been found \cite{TA}, illustrating
the distinction between entanglement and nonlocality first made by Werner
\cite{Werner}. The demonstration of ESD under noise in the case of multipartite states 
has been difficult because defining practical multipartite 
entanglement measures for the mixed states inevitably produced by 
such noise is highly nontrivial. %({\it cf.} \cite{BPRST}). 
Despite this difficulty, an phenomenon analogous to ESD
can more rigorously be studied in multipartite systems, 
namely, the effect of Bell Nonlocality Sudden Death (BNSD). 
The loss of nonlocal properties due to effects that are entirely local is the
most significant element of this, particularly in the case of multiple subsystems
where nonlocal behavior is not ``encoded'' in local states, 
as it can be in the case of pure bipartite two-level states. 
For example, the state entropy for the subsystems of a pair of
two-level systems determines the global properties of entanglement and 
nonlocality in the joint bipartite pure states ({\it cf.} \cite{BPRST}), whereas
for multipartite states, such a simple relationship no longer holds.

This effect was recently indicated by the demonstration \cite{JA08} that a 
tripartite system prepared in the W state initially violating the 
Mermin-Ardehali-Belinskii-Klyshko (MABK) inequality \cite{Mermin90,Ardehali92,BK93} 
fails to violate it at a later finite time in a local phase noise environment. The 
W class is a set of zero measure compared to the class of 
generic entangled pure states of three two-level systems \cite{DVC00}.
Here, a far stronger and more general result is obtained, 
namely, a definitive demonstration that the death 
of Bell nonlocality occurs suddenly in finite time in any system prepared 
in any one of the members of the generic class of 
tripartite-entangled pure states and subject to local phase noise alone, a result that 
requires the examination not only of the MABK inequality but of the full representative 
subset of the entire 256-element set of WWZB Bell-type inequalities \cite{WW01,ZB02} 
and the Svetlichny inequality for three two-level systems \cite{Svetlichny87}.
This result is demonstrated using an explicit and broadly applicable 
model which includes explicit time-dependence.

These results fall within the context of other recent results regarding 
decoherence of multipartite nonlocal quantum states. For example,
Sen(De), Sen, Wie$\acute{{\rm s}}$niak, Kaszlikowski, and 
$\dot{{\rm Z}}$ukowski \cite{SSW03} performed an analysis 
focused on nonlocality rather than entanglement; they considered 
the persistence of Bell-type nonlocality in multipartite GHZ and 
W states under multilocal phase noise and found that the 
nonlocality properties of W-type states were more robust against 
multilocal phase noise than those of the GHZ class.  Our results 
reinforce this latter observation by showing that, in the case of $n=3$ with
an explicit and physically motivated noise model, not only are the 
generalized GHZ state not robust, but they exhibit nonlocality sudden death.

In particular, we study the relatively small but illuminating case of triples 
of two-level systems in detail and demonstrate, for the first 
time in a situation where $n\geq 3$, that sudden death of multi-partite nonlocality 
occurs in a system for a range of state preparations due to such local phase noise 
alone.  Moreover, we show sudden death of two distinct types of nonlocal correlation: 
tripartite correlations associated with the inequality of Svetlichny \cite{Svetlichny87} and 
nonlocal correlations associated with the Werner and Wolf \cite{WW01} 
and $\dot{\rm Z}$ukowski  and Brukner \cite{ZB02} inequalities, which 
subsume the MABK form.  

We  thereby extend the study of Bell nonlocality 
sudden death in several ways. First, because previous sudden death results for 
tripartite states considered only correlations addressed by the MABK inequality, 
which is the representative of only one of the five distinct types of 
inequality of the full set of WWZB Bell-type inequalities \cite{JA08}, those
preliminary results concerned the sudden death of only one species 
of Bell-nonlocal correlations, whereas we here show the sudden failure to violate 
the entire 256-element set of WWZB Bell-type inequalities under local phase noise.  
That is, the sudden death of {\em all} species of Bell-nonlocal correlation in the 
presence of local dephasing noise alone is proven. Second, we 
demonstrate Bell nonlocality sudden death as captured by the Svetlichny inequality
for initially genuinely tripartite-entangled pure states of the generic class (GHZ-class) \cite{DVC00}.
Thus, we show that  Bell nonlocality sudden death occurs in this class of states in 
two distinct senses: there is the sudden loss of genuinely tripartite Bell nonlocality and 
of subsystem bipartite Bell nonlocality.  Finally, we explicitly confirm that  nonlocality death 
in the even-odd bipartite state-split of the system of three two-level systems occurs in 
precisely the same manner and timescale as that of genuinely tripartite Bell nonlocality
death.

The simple, pervasive character of the local phase noise considered here is noteworthy. 
Local phase noise appears in a broad range of physical situations and is of great concern, 
for example, in attempts to distribute quantum states, even in a very simple environment. 
That such a simple form of noise is unavoidable and can lead to the loss of Bell nonlocality 
for tripartite states is of great significance for entanglement distribution and quantum 
computing \cite{Unruh}, where entangled states of multiple two-level system appear in 
algorithms offering exponential speedups over classical computing and such states are 
used as encoding states \cite{Steane96}.

\section{BELL-TYPE NONLOCALITY IN VARIOUS CONTEXTS}

Svetlichny's Bell-type inequality \cite{Svetlichny87} distinguishes genuinely 
three-subsystem nonlocal correlations A-B-C of a system ABC composed of two-level 
subsystems A, B, and C, from those that can be described by a hybrid local-nonlocal 
model for a 1-2 subsystem A-BC (or B-AC or C-AB) bipartite split  
and furthermore, from ``convex sums'' of such hybrid local-nonlocal models.
In contrast, the Mermin-Ardehali-Belinskii-Klyshko (MABK) 
Bell-type inequality for three-component systems \cite{Mermin90,Ardehali92,BK93}, which 
has often been used in studies of nonlocality and was recently 
used to explore a precondition for Bell nonlocality sudden death 
\cite{JA08}, is incapable of addressing effects involving the element of 
genuinely  tripartite Bell-nonlocal correlation or loss thereof. Let us write
Svetlichny's inequality as
\begin{eqnarray}
|\mathcal{S}| \equiv |E(ABC) &+& E(ABC') + E(AB'C) + E(A'BC)  \\
 &-& E(A'B'C') - E(A'B'C) - E(A'BC') - E(AB'C')| \leq 4 \label{SvetlichnyInequality1} \nonumber , 
\end{eqnarray}
where $E(\cdots)$ denotes the expectation value of the measured outcomes in 
state-components $A, B$, and $C$, for example, a component of spin, primes 
denoting alternative directions of measurement.  When $|\mathcal{S}| > 4$, 
one has genuine tripartite Bell-nonlocal correlations, rather than simply 
bipartite correlations between subsystems within a tripartite system 
\cite{Cereceda02}; as expected, the maximum quantum value of  
$\max(|\mathcal{S}|) = 4\sqrt{2}$, compared to the algebraic maximum of 8, 
is attained only when the system is prepared in the maximally entangled ({\it cf.} \cite{CKW00})
GHZ state, $|{\rm GHZ}\rangle={1/\sqrt{2}}(|000\rangle+|111\rangle)$, the
representative of one of the two entanglement classes of tripartite pure 
states, the generic class \cite{DVC00}.

We refer to the following four distinct notions of Bell nonlocality in this paper.

\noindent i. {\it Generic Bell nonlocality} - The most general class of tripartite 
Bell nonlocality, for which Bell nonlocality of any type is present within the tripartite
system.  This class contains states in which Bell-locality and nonlocality may both be present
in subsystems or genuinely tripartite Bell non-locality may be present in the tripartite 
state.  All generic Bell nonlocality no longer exists when our state is describable 
using a local classical model, occurring when all of the WWZB inequalities are satisfied.

\noindent ii. {\it Genuinely tripartite Bell nonlocality} - Exists when the Svetlichny 
inequality for a tripartite state is violated, that is, when a hybrid local-nonlocal model 
cannot be used to describe the state.

\noindent iii. {\it Subsystem bipartite Bell nonlocality} - Refers to the nonlocality existing 
in a bipartite two-level system within a larger tripartite system, for example, the subset of 
bipartite two-level systems AB, BC, or AC within a tripartite system ABC.  Implicit in this 
definition is that each subsystem is nonlocally separated from the other subsystems.  This 
sort of nonlocality occurs when a {\it single} tripartite WWZB inequality is violated.  

\noindent iv. {\it Nonlocality of the even-odd bipartite state split} - Bell nonlocality for 
a bipartite two-level partition of the tripartite state.  Two of those four dimensions are 
within the Hilbert space of one two-level system and the other two dimensions are those of 
the remaining two-level system space. We can have, for example, two of four dimensions within 
the Hilbert space of subsystem A and the remaining two dimensions within the joint Hilbert 
space of B and C.  Regardless of how the two-level system pair split is made, one can analyze 
the corresponding Bell nonlocality properties using the CHSH inequality.  The development of
this type of Bell nonlocality, its significance, and its relation to the previous notions of 
Bell nonlocality is discussed in Sec. 4.

Despite the differences between the Svetlichny and the MABK inequalities, 
they can be related mathematically. 
Consider the following two pertinent instances of the MABK inequality.
\begin{eqnarray}
\left|\mathcal{M}\right|  &=& \left| E(ABC') + E(AB'C) + E(A'BC) - E(A'B'C') \right| \leq 2 \label{operatorM}\ , \\
\left|\mathcal{M}'\right| &=& \left| E(ABC) - E(AB'C') - E(A'BC') - E(A'B'C) \right| \leq 2 \label{operatorM'} \ ,
\end{eqnarray}
where $\mathcal{M}$ and $\mathcal{M}'$ are Bell-type operators, with differing arguments all 
of which appear in the single instance of the Svetlichny inequality above.  Either $|\mathcal{M}| > 2$ 
or $|\mathcal{M}'| > 2$ indicates the presence of Bell-nonlocal correlation via the MABK inequality, 
although this does not indicate genuine {\it tripartite} Bell-nonlocal correlation; 
tripartite Bell nonlocality is not guaranteed even when $\max(|\mathcal{M}|) = 
\max(|\mathcal{M}'|) = 4$, because these values can be achieved by convex
combinations of bipartite correlations alone. The left-hand-side of the Svetlichny 
inequality for genuine tripartite correlations is rather
\begin{eqnarray}
\left|\mathcal{S}\right| = \left|\mathcal{M} + \mathcal{M}'\right| \leq \left|\mathcal{M}\right| + \left|\mathcal{M}'\right|% \leq 4
\label{operatorS}\  .
\end{eqnarray}
For the state 
$|{\rm W}\rangle={1/\sqrt{3}}(|100\rangle+|010\rangle+|001\rangle)$, 
which is the representative of the other class of tripartite entangled states 
than that represented by $|{\rm GHZ}\rangle$, the maximum value attainable for the 
left-hand-side of the Svetlichny inequality is $\max(|\mathcal{S}|_{\rm W}) = 4.354>4$, 
which occurs when $\max(|\mathcal{M}|_{\rm W}) = \max(|\mathcal{M}^{'}|_{\rm W}) = 2.177$, 
which is inferior to the maximum quantum mechanical violation attained for the
GHZ-state, even though in this case the nonlocal correlations take the
form of convex combinations of bipartite Bell-nonlocal correlations.
Thus, the greatest possible extent of destruction of tripartite Bell nonlocality 
can be greater for states in the GHZ class.

The {\em necessary and sufficient} condition for the behavior 
of a system of three two-level subsystems to be describable by 
a fully Bell-local hidden-variables model, however, is provided jointly by the 
WWZB set of inequalities: {\em all elements} of the entire 
$256$-element set of WWZB Bell-type inequalities \cite{WW01,ZB02} 
must be satisfied for Bell locality and the violation of {\em even a single member} 
of the set of WWZB inequalities is sufficient for Bell nonlocality.  Therefore, in 
order to demonstrate the death of all Bell nonlocality in such a system due
to some physical influence, it is necessary for {\it all} members of this set of inequalities 
to become satisfied after at least one of them is not at some previous time, in
addition to the demonstration of the same for similar obeyance and violation  
of the Svetlichny inequality.  In Section IV below, this is shown to occur for states of the 
generic pure state entanglement class $|\Psi_3\rangle$, which is represented by 
the GHZ state. The WWZB inequalities are discussed in the next section, in its 
subsection B, after the pertinent noise model, states and notation is introduced in its 
subsection A.

\section{BELL NONLOCALITY SUDDEN DEATH IN THE TRIPARTITION}

Let us take the system of three two-level systems under study to be
prepared in the generic pure entanglement-class state \cite{AAL00}, 
\begin{equation}
\ket{\rm \Psi_3} = \bar{a}_{0}\ket{000} + \bar{a}_{4}\ket{100} + 
\bar{a}_{5}\ket{101} + \bar{a}_{6}\ket{110} + \bar{a}_{7}\ket{111}\ 
\end{equation} 
in $\mathcal{H}_{\rm ABC} = \mathcal{H}_{\rm A} \otimes \mathcal{H}_{\rm B} \otimes \mathcal{H}_{\rm C}$, 
where $\bar{a}_{i}\in\mathcal{C}$ and $\sum_{i}|\bar{a}_{i}|^2=1$, that is,
\begin{eqnarray}
\rho(0)=
\left(
\begin{array}{cccccccc}
 |\bar{a}_{0}|^2 \ & \ 0 \ & \ 0 \ & \ 0 \ & \ 
  \bar{a}_{0}\bar{a}_{4}^{*} \ & \ \bar{a}_{0}\bar{a}_{5}^{*} \ & \ \bar{a}_{0}\bar{a}_{6}^{*} \ &  \bar{a}_{0}\bar{a}_{7}^* \ \\
 0 & 0 & 0 & 0 & 0 & 0 & 0 & 0 \\
 0 & 0 & 0 & 0 & 0 & 0 & 0 & 0 \\
 0 & 0 & 0 & 0 & 0 & 0 & 0 & 0 \\
 \bar{a}_{4}\bar{a}_{0}^{*} & 0 & 0 & 0 & 
 |\bar{a}_{4}|^2 & \bar{a}_{4}\bar{a}_{5}^{*} & \bar{a}_{4}\bar{a}_{6}^{*} & \bar{a}_{4}\bar{a}_{7}^{*} \\
 \bar{a}_{5}\bar{a}_{0}^{*} & 0 & 0 & 0 & 
 \bar{a}_{5}\bar{a}_{4}^{*} & |\bar{a}_{5}|^2 & \bar{a}_{5}\bar{a}_{6}^{*} & \bar{a}_{5}\bar{a}_{7}^{*} \\
 \bar{a}_{6}\bar{a}_{0}^{*} & 0 & 0 & 0 & 
 \bar{a}_{6}\bar{a}_{4}^{*} & \bar{a}_{6}\bar{a}_{5}^{*} & |\bar{a}_{6}|^2 & \bar{a}_{6}\bar{a}_{7}^{*} \\
 \bar{a}_{7}\bar{a}_{0}^{*} & 0 & 0 & 0 & 
 \bar{a}_{7}\bar{a}_{4}^{*} & \bar{a}_{7}\bar{a}_{5}^{*} & \bar{a}_{7}\bar{a}_{6}^{*} & |\bar{a}_{7}|^2
\end{array}
\right) \ .
\end{eqnarray}

The tripartite generic state is analyzed because of its relations to the other pure
tripartite classes: GHZ, W, biseparable (B), and separable (S).  The generic state 
may be locally transformed with some finite probability into the GHZ class of states, 
which in turn may be converted stochastically by means positive-operators-valued 
measures (POVMs) into any of the other classes described by the following ordered 
relation \cite{DVC00}: $\rm{S} \subset \rm{B} \subset \rm{W} \subset \rm{GHZ}$.
The analysis of the generic tripartite state completes and extends the analysis of 
\cite{JA08}, where the phenomenon of Bell nonlocality sudden death was shown to exist 
in the W class. A feature that distinguishes the GHZ state from the W state is that 
the former is genuinely entangled at the tripartite level as opposed to the latter, 
which may be described by a convex sum of bipartite entangled states and is of measure zero.

The following results for the generic class of tripartite state 
apply immediately to the GHZ state itself, due to the measurement operators we have used 
that are composed of the tensored products of the Pauli matrices: $\sigma_{\rm X}$ and 
$\sigma_{\rm Y}$.  Furthermore, due to the fact that multi-local operations only cannot 
change the nonlocality properties of state, it is noteworthy that it suffices to use a 
specific GHZ state as representative of the GHZ class.  We have not assigned specific values 
to the coefficients in order to get the most general expressions for 
demonstration and clarity. However, we do assign them in specific instances to demonstrate, 
for example, the coefficients $\bar{a}_{0} = \bar{a}_{0}^{*} = \bar{a}_{7} = \bar{a}_{7}^{*} = 1/\sqrt{2}$
correspond  to maximum violation of the Svetlichny inequality and the longest timescale 
in which genuinely tripartite nonlocality is lost. \newline

Let the components of ABC be noninteracting and subject only to local external 
phase noise. The time-evolved state of an open quantum system 
under such external noise, written in the operator-sum representation, is
\begin{equation}
\rho\left(t\right) = \mathcal{E}\big(\rho\left(0\right)\big) =
\sum_{\mu}D_{\mu}\left(t\right)\rho\left(0\right)
D_{\mu}^{\dagger}\left(t\right) \ ,
\end{equation}
where the $\{D_\mu(t)\}$, with the index $\mu$ running
over all elements of the chosen operator-sum decomposition,
satisfy the completeness condition that guarantees that the evolution 
be trace-preserving
\cite{Kraus83}. For a collection of local noise sub-environments, 
noise operates locally on individual subsystems, that is, the
$D_{\mu}(t)$ are of the form $G_{k}(t)F_{j}(t)E_{i}(t)$. Hence,
\begin{eqnarray}
\rho\left(t\right) &=& \mathcal{E}\left(\rho\left(0\right)\right) =
\sum_{i = 1}^{2}\sum_{j = 1}^{2}\sum_{k = 1}^{2}
G_{k}\left(t\right)F_{j}\left(t\right)E_{i}\left(t\right)
\rho\left(0\right)
E_{i}^{\dagger}\left(t\right)F_{j}^{\dagger}\left(t\right)G_{k}^{\dagger}\left(t\right)\
.\label{krausSpecific}
\end{eqnarray}
In particular, let this local noise to be the basis-dependent
pure phase noise for which
\begin{eqnarray}
E_{1}(t) &=& {\rm diag}(1,\gamma_{\rm A}(t)) \otimes \mathbf{I}_{2} \otimes \mathbf{I}_{2} \ \ , \ \ 
E_{2}(t)  =  {\rm diag}(0,\omega_{\rm A}(t)) \otimes \mathbf{I}_{2} \otimes \mathbf{I}_{2} \ ,  \\
F_{1}(t) &=& \mathbf{I}_{2} \otimes {\rm diag}(1,\gamma_{\rm B}(t)) \otimes \mathbf{I}_{2} \ \ , \ \ 
F_{2}(t)  =  \mathbf{I}_{2} \otimes {\rm diag}(0,\omega_{\rm B}(t)) \otimes \mathbf{I}_{2} \ , \\
G_{1}(t) &=& \mathbf{I}_{2} \otimes \mathbf{I}_{2} \otimes {\rm diag}(1,\gamma_{\rm C}(t)) \ \ , \ \ 
G_{2}(t)  =  \mathbf{I}_{2} \otimes \mathbf{I}_{2} \otimes {\rm diag}(0,\omega_{\rm C}(t)) \ , \ 
\end{eqnarray}
$\gamma_{\rm A}\left(t\right) = \gamma_{\rm B}\left(t\right) = \gamma_{\rm C}\left(t\right) =
 \gamma\left(t\right) = e^{-\Gamma t},
 \omega_{\rm A}\left(t\right) = \omega_{\rm B}\left(t\right) = \omega_{\rm C}\left(t\right) =
 \omega\left(t\right) = \sqrt{1-\gamma^{2}(t)} = \sqrt{1-e^{- 2 \Gamma t}}$,
$\Gamma$ being the parameter describing the rate of local asymptotic 
dephasing taken to be that induced by all three sub-environments in their local subsystems:
The $\{E_{i}(t)\}$, $\{F_{j}(t)\}$, and $\{G_{k}(t)\}$ dephase the local 
state of each two-level subsystem individually at the same rate, $\Gamma$. For clarity,
the time-dependence of $\gamma(t)$'s are implicitly written from here on, 
particularly when displaying full density matrices.
This local phase noise appears in a broad range of physical situations and is of 
concern, for example, in attempts to distribute entanglement. That such a simple 
form of noise is unavoidable is of great significance for entanglement distribution 
and quantum computing. 

In the multi-local noise environment described above, 
for the composite system initially prepared at $t=0$ in 
%the generic entanglement class state 
$\rho(0)=\ket{\rm \Psi_{3}}\bra{\rm \Psi_3}$, the
solution of Eq. (8) at later time $t$ is
\begin{eqnarray}
{\hskip -28pt}\rho\left(t\right) {\hskip -6pt} = {\hskip -6pt}\left( {\hskip -6pt}
\begin{array}{cccccccc}
 |\bar{a}_{0}|^2 \ & \ 0 \ & \ 0 \ & \ 0 \ & \ 
  \bar{a}_{0}\bar{a}_{4}^{*}\gamma_{\rm A} \ & \ \bar{a}_{0}\bar{a}_{5}^{*}\gamma_{\rm A}\gamma_{\rm C} \ & \ 
  {\hskip -4pt}\bar{a}_{0}\bar{a}_{6}^{*}\gamma_{\rm A}\gamma_{\rm B} \ & 
  {\hskip -4pt} \bar{a}_{0}\bar{a}_{7}^{*}\gamma_{\rm A}\gamma_{\rm B}\gamma_{\rm C} \ \\
 0 & 0 & 0 & 0 & 0 & 0 & 0 & 0 \\
 0 & 0 & 0 & 0 & 0 & 0 & 0 & 0 \\
 0 & 0 & 0 & 0 & 0 & 0 & 0 & 0 \\
 \bar{a}_{4}\bar{a}_{0}^{*}\gamma_{\rm A} & 0 & 0 & 0 & 
 |\bar{a}_{4}|^2 & \bar{a}_{4}\bar{a}_{5}^{*}\gamma_{\rm C} & \bar{a}_{4}\bar{a}_{6}^{*}\gamma_{\rm B} & 
 \bar{a}_{4}\bar{a}_{7}^{*}\gamma_{\rm B}\gamma_{\rm C} \\
 \bar{a}_{5}\bar{a}_{0}^{*}\gamma_{\rm A}\gamma_{\rm C} & 0 & 0 & 0 & 
 \bar{a}_{5}\bar{a}_{4}^{*}\gamma_{\rm C} & |\bar{a}_{5}|^2 & 
 \bar{a}_{5}\bar{a}_{6}^{*}\gamma_{\rm B}\gamma_{\rm C} & \bar{a}_{5}\bar{a}_{7}^{*}\gamma_{\rm C} \\
 \bar{a}_{6}\bar{a}_{0}^{*}\gamma_{\rm A}\gamma_{\rm B} & 0 & 0 & 0 & 
 \bar{a}_{6}\bar{a}_{4}^{*}\gamma_{\rm B} & \bar{a}_{6}\bar{a}_{5}^{*}\gamma_{\rm B}\gamma_{\rm C} & 
 |\bar{a}_{6}|^2 & \bar{a}_{6}\bar{a}_{7}^{*}\gamma_{\rm C} \\
 \bar{a}_{7}\bar{a}_{0}^{*}\gamma_{\rm A}\gamma_{\rm C}\gamma_{\rm C} & 0 & 0 & 0 & 
 \bar{a}_{7}\bar{a}_{4}^{*}\gamma_{\rm B}\gamma_{\rm C} & \bar{a}_{7}\bar{a}_{5}^{*}\gamma_{\rm B} & 
 \bar{a}_{7}\bar{a}_{6}^{*}\gamma_{\rm C} & |\bar{a}_{7}|^2
\end{array}{\hskip  -10pt}
\right).
\end{eqnarray}
The off-diagonal elements of this matrix are seen to undergo asymptotic 
exponential decay with one of the rates $\Gamma$, $2\Gamma$, or $3\Gamma$. 
The full triple two-level system state, therefore, fully decoheres only 
in the infinite-time limit, because the off-diagonal dephasing factors 
$\gamma_{\rm A}$, $\gamma_{\rm B}$, and $\gamma_{C}$ only asymptotically
approach zero. Nonetheless, as we now demonstrate, the 
tripartite Bell-{\em nonlocality} of these states is entirely lost in a 
specific and finite time-scale.

The measurement operators $M_{K}$ and $M_{K}'$ 
of Eqs. (\ref{operatorM})-(\ref{operatorS}) in the 
Bell-type inequalities for $n=3$ correspond to measurements 
on each of the subsystems $K$  (A, B, or C), with the primed and 
unprimed terms denoting two different measurement directions 
for the corresponding party.  Defining 
$M_{\rm A} \equiv \sigma_{\rm y}$ and $M_{\rm A}' \equiv \sigma_{\rm x}$,
%which are the eigenoperators of the GHZ state, 
the measurement operator acting upon each successive subsystem is defined with 
respect to the first by a rotation by $\theta_{K}$:
\begin{eqnarray}
\left(
\begin{array}{c}
 M_{K} \\
 M_{K}'
\end{array}
\right) = R(\theta_{K})\left(
\begin{array}{c}
 M_{\rm A} \\
 M_{\rm A}' \end{array}
\right) \ , \ \ 
{\rm where} \ \  
R\left(\theta_{K}\right) = 
\left(
\begin{array}{cc}
 \cos \theta_{K} & -\sin \theta_{K}  \\
 \sin \theta_{K} & \ \ \cos \theta_{K}
\end{array}
\right) \ .
\end{eqnarray}
There are two such rotation 
angles $\theta_{\rm B}$ and $\theta_{\rm C}$ ($K = {\rm B}, {\rm C}$);
the corresponding measurement 
operators for two-level systems A, B, and C are 
\begin{eqnarray}
M_{\rm A} &=& \sigma_{\rm y} \otimes \mathbf{I}_{2} \otimes \mathbf{I}_{2} \ , \\
M_{\rm A}' &=& \sigma_{\rm x} \otimes \mathbf{I}_{2} \otimes \mathbf{I}_{2} \ , \\
M_{\rm B}  &=& \mathbf{I}_{2} \otimes
\left[\cos\left(\theta_{\rm B}\right)\sigma_{\rm y}-\sin\left(\theta_{\rm B}\right)\sigma_{\rm x}\right]
\otimes \mathbf{I}_{2} \ , \\
M_{\rm B}' &=& \mathbf{I}_{2} \otimes
\left[\sin\left(\theta_{\rm B}\right)\sigma_{\rm y}+\cos\left(\theta_{\rm B}\right)\sigma_{\rm x}\right]
\otimes \mathbf{I}_{2} \ , \\
M_{\rm C}  &=& \mathbf{I}_{2} \otimes \mathbf{I}_{2} \otimes
\left[\cos\left(\theta_{\rm C}\right)\sigma_{\rm y}-\sin\left(\theta_{\rm C}\right)\sigma_{\rm x}\right] \ , \\
M_{\rm C}' &=& \mathbf{I}_{2} \otimes \mathbf{I}_{2} \otimes
\left[\sin\left(\theta_{\rm C}\right)\sigma_{\rm y}+\cos\left(\theta_{\rm C}\right)\sigma_{\rm x}\right] \ .
\end{eqnarray}

\subsection{Svetlichny Inequality}

The Svetlichny operator appearing in Eq. (1) is, in terms of the measurement operators
introduced above,
\begin{eqnarray}
\mathcal{S} &=& 
M_{\rm A}M_{\rm B}M_{\rm C}  + M_{\rm A}M_{\rm B}M_{\rm C}' + 
M_{\rm A}M_{\rm B}'M_{\rm C} + M_{\rm A}'M_{\rm B}M_{\rm C} \nonumber \\
&-& M_{\rm A}'M_{\rm B}'M_{\rm C}' - M_{\rm A}'M_{\rm B}'M_{\rm C} - 
M_{\rm A}'M_{\rm B}M_{\rm C}' - M_{\rm A}M_{\rm B}'M_{\rm C}' \ . 
\end{eqnarray}
Recall that if $|\left\langle\mathcal{S}\right\rangle_{\rho(t)}| 
= {\rm tr}\left[ \mathcal{S}\rho(t) \right] > 4$, 
the state $\rho(t)$ is genuinely tripartite Bell nonlocal.  
In order to demonstrate tripartite Bell nonlocality sudden death in $\rho$ due to the effect
of external noise, we must show that both
$|\left\langle\mathcal{S}\right\rangle_{\rho(0)}| > 4$ 
and $|\left\langle\mathcal{S}\right\rangle_{\rho(t)}| \leq 4$
for some finite $t > 0$ under it. We now show that this 
indeed occurs for a system composed of three two-level subsystems prepared 
in generic state $\ket{\rm \Psi_{3}}$ under local phase noise 
described by the model of the previous section.  
Considering the complex coefficients $\bar{a}_{0}$ and $\bar{a}_{7}$ 
in polar forms $\bar{a}_{0} = |\bar{a}_{0}|e^{i\phi(\bar{a}_{0})}$ and 
$\bar{a}_{7} = |\bar{a}_{7}|e^{i\phi(\bar{a}_{7})}$,
let us write the relative phase angle between the amplitudes of the
amplitudes as $\alpha = \phi(\bar{a}_{0}) - \phi(\bar{a}_{7})$ and
$\theta_{\rm BC\alpha} = \theta_{\rm B} + \theta_{\rm C} + \alpha$.
Therefore,
\begin{eqnarray} {\hskip -26pt}
\left\langle \mathcal{S}\right\rangle_{\rho(t)} &=& {\rm tr}
\left[ \mathcal{S}\rho(t) \right] \nonumber \\
&=& {\rm tr} [(
M_{\rm A}M_{\rm B}M_{\rm C}  + M_{\rm A}M_{\rm B}M_{\rm C}' + 
M_{\rm A}M_{\rm B}'M_{\rm C} + M_{\rm A}'M_{\rm B}M_{\rm C} \nonumber\\
& & \ \ \ \ \ \ \ \ \ \ \ \ \ \ \ \ \ \ \ \ - M_{\rm A}'M_{\rm B}'M_{\rm C}' - M_{\rm A}'M_{\rm B}'M_{\rm C} - 
M_{\rm A}'M_{\rm B}M_{\rm C}' - M_{\rm A}M_{\rm B}'M_{\rm C}'
)\rho(t)] \nonumber \\
&=& (4 + 4i)\gamma_{\rm A}\gamma_{\rm B}\gamma_{\rm C}
\left[
(i\bar{a}_{7}\bar{a}_{0}^{*} -   \bar{a}_{0}\bar{a}_{7}^{*})\cos(\theta_{\rm B} + \theta_{\rm C}) +
( \bar{a}_{7}\bar{a}_{0}^{*} - i \bar{a}_{0}\bar{a}_{7}^{*})\sin(\theta_{\rm B} + \theta_{\rm C})
\right] \nonumber \\
&=& 8 \gamma_{\rm A}\gamma_{\rm B}\gamma_{\rm C}
\left|\bar{a}_{0}\right|\left|\bar{a}_{7}\right| 
\left[ 
\cos\left(\theta_{\rm BC\alpha}\right) - 
\sin\left(\theta_{\rm BC\alpha}\right)
\right] \ .
\end{eqnarray}

The Svetlichny inequality is violated whenever 
$|\left\langle \mathcal{S}\right\rangle_{\rho(t)}| > 4$, which is seen to occur for any state 
$|\Psi_3\rangle$ for which $|\bar{a}_{0}||\bar{a}_{7}| > 1 /(2\sqrt{2})$,
the maximal violation for each state occurring at $\theta_{\rm BC\alpha} = -\pi / 4 \ , 3\pi / 4$ 
and $t = 0$, at which time $\gamma_{\rm A} = \gamma_{\rm B} = \gamma_{\rm C} = 1$.
The maximum quantum mechanically allowed value, 
$|\left\langle \mathcal{S}\right\rangle_{\rho(t)}| = 4\sqrt{2}$, 
is attained by elements of the generic class $\ket{\Psi_{3}}$ 
for which $|\bar{a}_{0}| = |\bar{a}_{7}| = 1 / \sqrt{2}$, for example, the standard GHZ state with
$\theta_{\rm BC}=-\pi / 4$ at $t=0$.
Furthermore, recalling that $\gamma_{\rm A} = \gamma_{\rm B} = \gamma_{\rm C} = e^{-\Gamma t}$
and assuming a natural local decoherence rate of $\Gamma = 1$, one sees that
the maximum value of the left-hand-side of the Svetlichny inequality for these 
initially tripartite Bell-nonlocal states evolves according to
$|\left\langle\mathcal{S}\right\rangle_{\rho(t)}| = 
8\sqrt{2}|\bar{a}_{0}||\bar{a}_{7}|e^{-3 \Gamma t}$, and so
approaches the critical value
$|\left\langle\mathcal{S}\right\rangle_{\rho(t_{3}^{*})}| = 4$ 
in the finite timescale
\begin{equation}
t^{*}_{3} = \frac{\ln(2\sqrt{2}|\bar{a}_{0}||\bar{a}_{7}|)}{3 \Gamma} \ .
\end{equation}

Thus, for example, when the system is initially prepared in the standard GHZ state, we find
$t^{*}_{3} = {\ln(\sqrt{2})}/{3 \Gamma}.$
For any initial preparation of the pure generic class, genuine tripartite Bell nonlocality is lost from  $t_{3}^{*}$ onward.

Before proceeding, one should note note that there exist ``two'' Svetlichny inequalities.  
The second Svetlichny inequality, denoted by $\mathcal{S'}$ is given by
\begin{eqnarray}
|\mathcal{S'}| \equiv |E(ABC) &-& E(ABC') - E(AB'C) - E(A'BC)  \\
 &+& E(A'B'C') - E(A'B'C) - E(A'BC') - E(AB'C')| \leq 4 \label{SvetlichnyInequality2} \nonumber . 
\end{eqnarray}
In particular, note that a minus sign appears in front of $E(ABC')$. (Also note that a typographical
error was made in front of that term in Eq. 6 of the published version of Svetlichny's original 
paper of 1987 \cite{Svetlichny87}, which was pointed out in footnote 9 of a later paper \cite{SS02}.)
In the current analysis, $\mathcal{S'} = - \mathcal{S}$ upon the substitution 
$\theta_{\rm BC\alpha} \rightarrow -\theta_{\rm BC\alpha}$; because only the maximum 
magnitude of the Svetlichny expression is relevant in this analysis, one gets 
similar results for $\mathcal{S'}$, so that here it is only necessary to refer to $\mathcal{S}$.

\subsection{WWZB Inequality}

Werner and Wolf \cite{WW01} and Zukowski and Brukner \cite{ZB02} 
have derived a set of $2^{2^n}$ Bell-type inequalities the conjunction 
of the truth values of which is a necessary and sufficient condition 
for a system composed of $n$ two-level subsystems to be describable 
by a fully local hidden-variables model.  For $n = 3$, there are 256 of these 
inequalities, which fall into five classes with elements forming 
subsets related by symmetries under (1) changing the labels of the 
measured observables at each site, (2) changing the names of the 
measurement outcomes, or (3) permuting subsystems. 
The behavior of a single element of each class is identical to that of 
all members of that class, as explicitly shown in the 
appendix of \cite{WW01}.  As a result, one need consider only one
inequality from each of the five distinct classes, for example, those 
with left-hand-sides with Bell-type operators of the forms

\begin{eqnarray}
{\rm (P1)}\ \ \ \ \ \ \mathcal{B}_{{\rm P}1}&=&
2 M_{\rm A}M_{\rm B}M_{\rm C} \ , \nonumber \\
{\rm (P2)}\ \ \ \ \ \ \mathcal{B}_{{\rm P}2}&=& \frac{1}{2}
(- M_{\rm A}M_{\rm B}M_{\rm C}   + M_{\rm A}M_{\rm B}M_{\rm C}'
+ M_{\rm A}M_{\rm B}'M_{\rm C}  + M_{\rm A}M_{\rm B}'M_{\rm C}' \nonumber \\
&+& M_{\rm A}'M_{\rm B}M_{\rm C}  + M_{\rm A}'M_{\rm B}M_{\rm C}' 
+ M_{\rm A}'M_{\rm B}'M_{\rm C} + M_{\rm A}'M_{\rm B}'M_{\rm C}'
)\ , \nonumber \\
{\rm (P3)}\ \ \ \ \ \ \mathcal{B}_{{\rm P}3}&=&
[M_{\rm A}(M_{\rm B} + M_{\rm B}') + M_{\rm A}'(M_{\rm B} - M_{\rm B}') ]M_{\rm C}\ , \nonumber \\
%[ a_{1} ( b_{1} + b_{2} ) + a_{2} ( b_{1} - b_{2} ) ] c_{1}\ , \nonumber \\
{\rm (P4)}\ \ \ \ \ \ \mathcal{B}_{{\rm P}4}&=& 
M_{\rm A}M_{\rm B}(M_{\rm C} + M_{\rm C}') - M_{\rm A}'M_{\rm B}'(M_{\rm C} - M_{\rm C}')\ , \nonumber \\
%a_{1}b_{1}( c_{1} + c_{2} ) - a_{2}b_{2}( c_{1} - c_{2} )\ ,  \nonumber \\
{\rm (P5)}\ \ \ \ \ \ \mathcal{B}_{{\rm P}5}&=& 
M_{\rm A}M_{\rm B}M_{\rm C}' + M_{\rm A}M_{\rm B}'M_{\rm C} +
M_{\rm A}'M_{\rm B}M_{\rm C} - M_{\rm A}'M_{\rm B}'M_{\rm C}' \ ,
\end{eqnarray}
which we consider here. For the entire class of local hidden-variables models, the 
corresponding Bell-type inequalities are 
$|\left\langle \mathcal{B}_{{\rm P} I} \right\rangle_{\rho}| \leq 2$
(for $I$ = 1, 2, 3, 4, 5). In order to show definitively that Bell nonlocality 
sudden death occurs in a system at $t^{*}$ for a class of state 
preparations, one must demonstrate both that (i) these system 
states are initially incapable of description by a local 
hidden-variables model at $t=0$, that is, that at least one of 
the ${\rm P} I > 2$ (for $I$ = 1, 2, 3, 4, 5), and (ii) they 
are describable by a hidden-variables model at some later time 
$t^{*} < \infty$, that is, 
$|\left\langle \mathcal{B}_{{\rm P} I} \right\rangle_{\rho}| \leq 2$
for all $I$ = 1, 2, 3, 4, 5 at $t^*$.

We first show (i), in particular, that at time $t=0$ the inequality of form P5 is violated by 
the same range of generic entanglement class pure states as considered above, 
and therefore that the system is not describable by an entirely local hidden-variables model---as 
opposed to local-nonlocal hybrid model, as pertained in Subsection IIA.
The expectation value of the $\mathcal{B}_{\rm P5}$ operator for the
state under the influence of multi-local noise on the composite system of
three two-level systems ABC initially prepared in the GHZ-class pure state is
\begin{eqnarray} {\hskip -26pt}
\left\langle \mathcal{B}_{\rm P5} \right\rangle_{\rho(t)} &=&
{\rm tr} \left[\mathcal{B}_{\rm P5}\rho(t)\right] \nonumber \\
&=& {\rm tr} \left[ 
\big(
M_{\rm A}M_{\rm B}M_{\rm C}' + M_{\rm A}M_{\rm B}'M_{\rm C} +
M_{\rm A}'M_{\rm B}M_{\rm C} - M_{\rm A}'M_{\rm B}'M_{\rm C}'
\big)
\rho(t) \right] \nonumber \\
&=& 4 \gamma_{\rm A}\gamma_{\rm B}\gamma_{\rm C} 
\left[ (\bar{a}_{0}\bar{a}_{7}^{*} + \bar{a}_{7}\bar{a}_{0}^{*}) 
\cos(\theta_{\rm B} + \theta_{\rm C}) -
i (\bar{a}_{0}\bar{a}_{7}^{*} - \bar{a}_{7}\bar{a}_{0}^{*})\sin(\theta_{\rm B} + 
\theta_{\rm C})\right] \nonumber \\
&=& 8 \gamma_{\rm A}\gamma_{\rm B}\gamma_{\rm C}
\left|\bar{a}_{0}\right|\left|\bar{a}_{7}\right| 
\sin\left(\theta_{\rm BC\alpha}\right)\ .
\end{eqnarray}
 Taking 
$\gamma_{\rm A} = \gamma_{\rm B} = \gamma_{\rm C} = e^{-\Gamma t}$
as before, the left-hand-side of this form of inequality evolves as
$|\left\langle \mathcal{B}_{\rm P5} \right\rangle_{\rho(t)}| = 8|\bar{a}_{0}||\bar{a}_{7}|e^{-3 \Gamma t}$ 
approaching the critical value 
$|\left\langle \mathcal{B}_{\rm P5} \right\rangle_{\rho(t^{*})}| = 2$ from above on a timescale
\begin{equation}
t^{*} = \frac{\ln(4|\bar{a}_{0}||\bar{a}_{7}|)}{3 \Gamma} \ .
\end{equation}
For example, the maximum value $|\left\langle \mathcal{B}_{\rm P5} \right\rangle_{\rho(t)}| = 4$
for initial Bell nonlocality occurs at $t = 0$ (when $\gamma_{\rm A} = \gamma_{\rm B} = \gamma_{\rm C} = 1$),
for $|\bar{a}_{0}| = |\bar{a}_{7}| = 1 / \sqrt{2}$, that is,
in the standard GHZ state (for which $\alpha=0$) and when the trigonometric term takes its maximum value, 
$\sin(\theta_{\rm BC\alpha}) = 1$, that is,
when $\theta_{\rm BC} = \pi / 2$; the critical time is then $t^{*}=\ln(2)/{3\Gamma}<\infty$.

We now show (ii), that is, that all the remaining inequalities, given this set of
initial state preparations, are later satisfied in the time scale $t^{*}$, 
so that the condition for a local hidden-variables model to suffice to explain the
resulting correlations is satisfied in it.  This occurs when the 
absolute value of the left-hand-side of the following expressions are 
less than or equal to the value two. Let us evaluate the operator expectation 
values $\left\langle \mathcal{B}_{{\rm P}I} \right\rangle_{\rho(t)}$, 
for each remaining inequality for $I = 1, 2, 3, 4$ in turn. 
\begin{eqnarray} {\hskip -26pt}
\left\langle \mathcal{B}_{\rm P1}\right\rangle_{\rho(t)} 
&=& {\rm tr}\left[\mathcal{B}_{P1}\rho(t)\right] \nonumber \\
&=& {\rm tr}\left[(
2 M_{\rm A}M_{\rm B}M_{\rm C}
)\rho(t)\right] \nonumber \\
&=& 2\gamma_{\rm A}\gamma_{\rm B}\gamma_{\rm C} 
\left[
(\bar{a}_{0}\bar{a}_{7}^{*} + \bar{a}_{7}\bar{a}_{0}^{*})
\sin(\theta_{\rm B} + \theta_{\rm C}) +
i (\bar{a}_{7}\bar{a}_{0}^{*} - \bar{a}_{0}\bar{a}_{7}^{*})
\cos(\theta_{\rm B} + \theta_{\rm C})
\right] \nonumber \\
&=& 4\gamma_{\rm A}\gamma_{\rm B}\gamma_{\rm C} 
|\bar{a}_{0}||\bar{a}_{7}|\sin(\theta_{\rm BC\alpha})\ .
\end{eqnarray}
One immediately sees that
$\max(|\left\langle \mathcal{B}_{\rm P1}\right\rangle_{\rho(t)}|) 
= 4|\bar{a}_{0}||\bar{a}_{7}|e^{-3\Gamma t} \leq 2$ for all times $t>0$ and over the full 
range of values of $\theta_{\rm BC\alpha}$, because 
for the states of interest $1/2\geq|\bar{a}_{0}||\bar{a}_{7}|> 1/4$, where
the upper bound 1/2 represents a maximally entangled state and the lower bound 1/4
represents a maximally mixed state.

For the inequality of form P2, one finds
\begin{eqnarray}
\left\langle \mathcal{B}_{\rm P2}\right\rangle_{\rho(t)} 
&=& {\rm tr}\left[\mathcal{B}_{\rm P2}\rho(t)\right] \nonumber \\
&=& {\rm tr}
\Big[\frac{1}{2}
\big(- M_{\rm A}M_{\rm B}M_{\rm C}   + M_{\rm A}M_{\rm B}M_{\rm C}'
+ M_{\rm A}M_{\rm B}'M_{\rm C}  + M_{\rm A}M_{\rm B}'M_{\rm C}' \nonumber \\
&&{\hskip 26pt}+ M_{\rm A}'M_{\rm B}M_{\rm C}  + M_{\rm A}'M_{\rm B}M_{\rm C}' 
+ M_{\rm A}'M_{\rm B}'M_{\rm C} + M_{\rm A}'M_{\rm B}'M_{\rm C}'
\big)\rho(t)\Big] \nonumber \\
&=& - \left(1 + i \right)\gamma_{\rm A}\gamma_{\rm B}\gamma_{\rm C}
[
\left(
[2 + i]   \bar{a}_{7}\bar{a}_{0}^{*} -
[1 + 2 i] \bar{a}_{0}\bar{a}_{7}^{*}
\right)\cos(\theta_{\rm B}+\theta_{\rm C}) \nonumber \\
& & \ \ \ \ \ \ \ \ \ \ \ \ \ \ \ \ \ \ \ \ \ +
\left(
[1 - 2i] \bar{a}_{7}\bar{a}_{0}^{*} +
[2 - i]  \bar{a}_{0}\bar{a}_{7}^{*}
\right)\sin(\theta_{\rm B}+\theta_{\rm C})
] \nonumber \\
&=& 2 \gamma_{\rm A}\gamma_{\rm B}\gamma_{\rm C}
|\bar{a}_{0}||\bar{a}_{7}|
\left[
3\sin(\theta_{\rm BC\alpha}) + \cos(\theta_{\rm BC\alpha})
\right] \ .
\end{eqnarray}
One sees that for $t \geq t^{*}$, 
$\left\langle \mathcal{B}_{\rm P2}\right\rangle_{\rho(t)} < 2$ for 
all choices of $\theta_{\rm BC\alpha}$: in that range the maximum with respect to
$\theta_{{\rm BC}\alpha}$ of $|\left\langle \mathcal{B}_{\rm P2}\right\rangle_{\rho(t^{*})}| =
 8|\bar{a}_{0}||\bar{a}_{7}| e^{- 3\Gamma t^{*}} <2$, because the trigonometric factor is strictly bounded by 4.

For the inequality of form P3, one finds
\begin{eqnarray}{\hskip -26pt}
\left\langle \mathcal{B}_{\rm P3}\right\rangle_{\rho(t)} 
&=& {\rm tr}\left[\mathcal{B}_{\rm P3}\rho(t)\right] \nonumber \\
&=& {\rm tr}\big[ \big(
M_{\rm A}M_{\rm B}M_{\rm C} + M_{\rm A}M_{\rm B}M_{\rm C}' +
M_{\rm A}'M_{\rm B}'M_{\rm C}' - M_{\rm A}'M_{\rm B}'M_{\rm C}
\big) \rho(t) \big]\nonumber \\
&=&
2\gamma_{\rm A}\gamma_{\rm B}\gamma_{\rm C}\left( 1 + i \right)
\left[
\left( i \bar{a}_{7}\bar{a}_{0}^{*} - \bar{a}_{0}\bar{a}_{7}^{*} \right)
\cos(\theta_{\rm B} + \theta_{\rm C})+
\left( \bar{a}_{7}\bar{a}_{0}^{*} - i \bar{a}_{0}\bar{a}_{7}^{*} \right)
\sin(\theta_{\rm B} + \theta_{\rm C})
\right]\nonumber \\
&=&
4\gamma_{\rm A}\gamma_{\rm B}\gamma_{\rm C}|\bar{a}_{0}||\bar{a}_{7}|
[\cos(\theta_{\rm BC\alpha}) - \sin(\theta_{\rm BC\alpha})] \ . 
\end{eqnarray}
The maximum with respect to $\theta_{{\rm BC}\alpha}$ is
$|\left\langle \mathcal{B}_{\rm P3}\right\rangle_{\rho(0)}| = 
4\sqrt{2}|\bar{a}_{0}||\bar{a}_{7}|$, which occurs for
$\theta_{\rm BC\alpha} = -\pi / 4, 3\pi/4$. At $t^{*}$, one has
$|\left\langle \mathcal{B}_{\rm P3}\right\rangle_{\rho(t^{*})}| 
= 4\sqrt{2} |\bar{a}_{0}||\bar{a}_{7}|e^{- 3\Gamma t^{*}} = 2\sqrt{2}|\bar{a}_{0}||\bar{a}_{7}|\leq 2$, 
and similarly for all later times for these optimal angles

Finally, for the remaining form, P4, one finds
\begin{eqnarray}{\hskip -26pt}
\left\langle \mathcal{B}_{\rm P4}\right\rangle_{\rho(t)} 
&=& {\rm tr}\left[\mathcal{B}_{\rm P4}\rho(t)\right] \nonumber \\
&=& {\rm tr}\big[\big(
M_{\rm A}M_{\rm B}M_{\rm C} + M_{\rm A}M_{\rm B}M_{\rm C}' + 
M_{\rm A}'M_{\rm B}'M_{\rm C}' - M_{\rm A}'M_{\rm B}'M_{\rm C}
\big)\rho(t)\big] \nonumber \\
&=& 2 \left[(\bar{a}_{0}\bar{a}_{7}^{*} + \bar{a}_{7}\bar{a}_{0}^{*})
\sin(\theta_{\rm B} + \theta_{\rm C}) +
i (\bar{a}_{7}\bar{a}_{0}^{*} - \bar{a}_{0}\bar{a}_{7}^{*})
\cos(\theta_{\rm B} + \theta_{\rm C})
\right]\gamma_{\rm A}\gamma_{\rm B}\gamma_{\rm C} \nonumber \\
&=& 4\gamma_{\rm A}\gamma_{\rm B}\gamma_{\rm C}|\bar{a}_{0}||\bar{a}_{7}|
\sin(\theta_{BC\alpha})\ .
\end{eqnarray}
\noindent One sees immediately, as in case P1, that 
$|\left\langle \mathcal{B}_{\rm P4}\right\rangle_{\rho(t)}| =
4|\bar{a}_{0}||\bar{a}_{7}| e^{-3\Gamma t} \leq 2$, for all times $t$ and for all values of 
$\theta_{\rm BC\alpha}$.

Thus, in the timescale 
$t^{*} = \ln(4|\bar{a}_{0}||\bar{a}_{7}|)/3\Gamma$, all the WWZB inequalities are 
satisfied for all measurement angles and all initially Bell-nonlocal generic pure-state entanglement class preparations $|\Psi_3\rangle$. Therefore,
the composite quantum system has entirely and irreversibly 
lost its Bell nonlocality in finite time under the influence only of local phase noise.

\section{BELL NONLOCALITY SUDDEN DEATH IN THE BIPARTITIONS}

The destruction of genuine tripartite Bell nonlocality in finite time for states
of three two-level systems was demonstrated in Section IIIA above using the Svetlichny 
inequality. In a three-component system, the loss of genuine tripartite Bell 
nonlocality entails the loss in the same system considered as composed of two 
subsystems of bipartite Bell nonlocality, one subsystem ({\it e.g.}, A) being one of the two-level 
systems alone and the other being the subsystem constituted by the remaining 
pair of two-level systems ({\it e.g.}, BC). We now verify that this is indeed the case, 
by considering the remaining two systems as a single unit in a bipartition of the system. 
In particular, we show that bipartite Bell nonlocality sudden death occurs, by using 
the CHSH inequality, in exactly the same time scale found when using the Svetlichny 
inequality.

Without loss of generality, because in our model local phase noise affects each
subsystem in exactly the same way, we take the solo two-level system to
be subsystem A and the remaining subsystems,  B and C, to jointly form 
subsystem BC with states lying in a four-dimensional Hilbert 
space $\mathcal{H}_{\rm BC}=\mathcal{H}_{\rm B} \otimes \mathcal{H}_{\rm C}$.  
The maximally Bell-nonlocal state in this bipartite splitting of the system corresponds
to the GHZ state, as can be seen by noting that with 
$\ket{\bar{0}}\equiv\ket{00}=(1, 0, 0, 0)^{\rm T}\in\mathcal{H}_{\rm BC}$ and
$\ket{\bar{1}}\equiv\ket{11}=(0, 0, 0, 1)^{\rm T}\in\mathcal{H}_{\rm BC}$,
respectively, the GHZ state is formally similar to the Bell state $\ket{\Phi^+}$, in that
$\ket{\rm GHZ}={1/\sqrt{2}}(\ket{0\bar{0}}+\ket{1\bar{1}})$. 
This decomposition is of the Schmidt form, which naturally exposes nonlocal
correlations, and shows how one can construct the 
CHSH spin-measurement operators in the two-dimensional subspace of 
$\mathcal{H}_{\rm BC}$, in terms of which the measurement outcomes on the quantum states 
are written when evaluating the inequality. %appropriate to the bipartition. 
In particular, writing $\tau=\ket{\bar{1}}\bra{\bar{0}} = \ket{11}\bra{00}$ and 
$\tau^{\dagger} = \ket{\bar{0}}\bra{\bar{1}}=\ket{00}\bra{11}$,
the Pauli-operator analogues are $\tau_{1} = \tau + \tau^{\dagger}, \tau_{2} = i\tau - i\tau^{\dagger},
\tau_{3} = \tau^{\dagger}\tau - \tau \tau^{\dagger}, \mathbf{I}_{\tau}=\tau^{\dagger}\tau + \tau\tau^{\dagger},$ where 
$\tau_{1} = \sigma_{\rm y} \otimes \sigma_{\rm x}$ and 
$\tau_{2} = \sigma_{\rm x} \otimes \sigma_{\rm x}$: 
One sees that $\tau_{1}$ and $\tau_{2}$ act analogously on $\ket{\bar{0}}$ and
$\ket{\bar{1}}$ as $\sigma_{\rm x}$ and $\sigma_{\rm y}$ act 
on the natural basis states of two-dimensional Hilbert space: 
$\tau_1\ket{\bar{0}}=\tau_{1}\ket{00}=\ket{11}=\ket{\bar{1}}\ , \tau_{1}\ket{\bar{1}}=\tau_1\ket{11}=\ket{00}=\ket{\bar{0}}, 
\tau_{2}\ket{\bar{0}}=\tau_2\ket{00}=i\ket{11}=i\ket{\bar{1}},$ and 
$\tau_{2}\ket{\bar{1}}=\tau_2\ket{11}=-i\ket{00}=-i\ket{\bar{0}}$,
as required.

The measurement generators appearing in the Bell operator
of the CHSH inequality, therefore, for the first subsystem are the usual ones and, for
the larger, second subsystem are
\begin{eqnarray}
\left(
\begin{array}{c}
 \bar{M}_{\rm BC} \\
 \bar{M}_{\rm BC}'
\end{array}
\right) = R(\theta_{\rm BC})\left(
\begin{array}{c}
 \bar{M}_{\rm A}\\
 \bar{M}_{\rm A}'
 %\tau_{2} \\
 %\tau_{1} 
 \end{array}
\right) \ , \ {\rm with} \ 
R\left(\theta_{\rm BC}\right) = 
\left(
\begin{array}{cc}
 \cos \theta_{\rm BC} & -\sin \theta_{\rm BC}  \\
 \sin \theta_{\rm BC} &  \cos \theta_{\rm BC} 
 \end{array}
\right) , 
\end{eqnarray}
that is, 
\begin{eqnarray}
\bar{M}_{\rm A}   &=& \sigma_{\rm y} \otimes \mathbf{I}_4 \ , \\
\bar{M}^{'}_{\rm A}  &=& \sigma_{\rm x} \otimes \mathbf{I}_4 \ , 
\end{eqnarray}
and
\begin{eqnarray}
\bar{M}_{\rm BC}  &=& \mathbf{I}_2 \otimes
\left[\cos\left(\theta_{\rm BC}\right)\tau_{2} -
\sin\left(\theta_{\rm BC}\right)\tau_{1}\right] \ , \\
\bar{M}^{'}_{\rm BC} &=& \mathbf{I}_2 \otimes
\left[\sin\left(\theta_{\rm BC}\right)\tau_{2} +
\cos\left(\theta_{\rm BC}\right)\tau_{1}\right] \ .
\end{eqnarray}
In terms of these measurement operators, the appropriate Bell-CHSH operator is then
\begin{equation}
\mathcal{B}_{\rm CHSH} = 
\bar{M}_{\rm A}\bar{M}_{\rm BC}     + \bar{M}_{\rm A}\bar{M}^{'}_{\rm BC} + 
\bar{M}^{'}_{\rm A}\bar{M}_{\rm BC} - \bar{M}^{'}_{\rm A}\bar{M}^{'}_{\rm BC} \ .
\end{equation}
Writing $\bar{\theta}_{\rm BC\alpha} = \theta_{\rm BC} + \alpha$, 
the Bell-operator expectation value for state $\rho(t)$ is 
\begin{eqnarray}{\hskip -26pt}
\left\langle \mathcal{B}_{\rm CHSH} \right\rangle_{\rho(t)} &=& 
{\rm tr} \left[\mathcal{B}_{\rm CHSH} \rho(t) \right] \nonumber \\  
&=& {\rm tr} \left[ \left(
\bar{M}_{\rm A}\bar{M}_{\rm BC}     + \bar{M}_{\rm A}\bar{M}^{'}_{\rm BC} + 
\bar{M}^{'}_{\rm A}\bar{M}_{\rm BC} - \bar{M}^{'}_{\rm A}\bar{M}^{'}_{\rm BC}
\right) \rho(t)\right]\nonumber \\
&=&  -(2 + 2i)\gamma_{\rm A}\gamma_{\rm B}\gamma_{\rm C}
 \left[ 
 (   \bar{a}_{7}\bar{a}_{0}^{*} - i \bar{a}_{0}\bar{a}_{7}^{*} )\cos(\theta_{\rm BC}) + 
 (-i \bar{a}_{7}\bar{a}_{0}^{*} +   \bar{a}_{0}\bar{a}_{7}^{*} )\cos(\theta_{\rm BC})
 \right]
 \nonumber \\
&=& 4 \gamma_{\rm A}\gamma_{\rm B}\gamma_{\rm C} |\bar{a}_{0}||\bar{a}_{7}|
\left[ \cos(\bar{\theta}_{\rm BC\alpha}) - \sin(\bar{\theta}_{\rm BC\alpha}) \right]\ .
\end{eqnarray}
\noindent Recall that
$| \left\langle \mathcal{B}_{\rm CHSH} \right\rangle_{\rho(t)} | \leq 2$
holds for all local hidden-variables models and that $2\sqrt{2}$ is the Tsirel'son
bound \cite{Tsirelson80,Tsirelson87}, the maximum violation attainable by 
quantum mechanical states. Whenever
$| \left\langle \mathcal{B}_{\rm CHSH} \right\rangle_{\rho(t)} |>2$
the system in $\rho$ exhibits Bell nonlocality. 
Before local phase decoherence begins at $t = 0$, 
the left-hand-side of the CHSH inequality is maximized when
$|\cos(\bar{\theta}_{\rm BC\alpha}) - \sin(\bar{\theta}_{\rm BC\alpha})| = \sqrt{2}$,
that is, when $\bar{\theta}_{\rm BC\alpha} = -\pi /4, 3\pi/4$, 
and $|\bar{a}_{0}|=|\bar{a}_{7}|=1/\sqrt{2}$, 
so that $|\left\langle \mathcal{B}_{\rm CHSH}\right\rangle_{\rho(t)}| = 2\sqrt{2}$.
After the local dephasing noise has begun acting, one finds that 
$\left\langle \mathcal{B}_{\rm CHSH} \right\rangle_{\rho(t)} = 2$
in the timescale 
\begin{equation}
t_{\rm 2}^{*} =\frac{\ln(2\sqrt{2}|\bar{a}_{0}||\bar{a}_{7}|)}{3 \Gamma}\ .
\end{equation}

The extent of inequality violation is thus seen to evolve in time in exactly the 
same manner as the violation of Svetlichny inequality.
In particular, one sees that Bell nonlocality sudden death occurs in
precisely the same timescale in these alternative perspectives on the same process, 
that is,  $t_{\rm 2}^{*} = t_{3}^{*}$, as it should.

%%%%%%%%%%%%%%%%%%%%%%%%%%%%%%%%%%%%%%%%%%%%%%%%%%%%%%%%%%%%%%%%%%%%%%%%%%%%%%%%%%%%%%%%%%%%%

\section{Conclusion}

We have shown that local phase noise that is capable of eliminating 
all state coherence only in the infinite-time limit is 
nonetheless capable of eliminating Bell nonlocality in finite time, for
three-component systems prepared in the generic entanglement class of 
tripartite states for all preparations in which they are initially Bell nonlocal. 
It is noteworthy that the noise acting on the initially 
entangled states is merely local, whereas the central characteristic of entanglement 
and Bell-type inequality violation is {\em nonlocality}.

This Bell nonlocality sudden death was examined in both of its aspects. 
One is the certain sudden death of all Bell-nonlocal correlations irreducible to convex sums of 
internal bipartite correlations in such states, exhibited by the sudden failure to violate 
Svetlichny's inequality. The other is the certain sudden death of Bell-nonlocal 
correlations reducible to such convex combinations of bipartite correlations, in 
that the three subsystems suddenly become jointly describable 
by a fully local hidden-variables model, as exhibited by their suddenly obeying the entire 
set of Werner--Wolf--${\rm\dot{Z}}$ukowski--Brukner inequalities. 
The results were also shown to accord with the behavior of correlations under 
bi-partitioning of the system. The loss of nonlocal properties due to effects that are 
entirely local is the most significant element of this, particularly in the case of 
multiple subsystems where nonlocal behavior is not ``encoded'' in local states,
as it can sometimes be in the case of bipartite two-level states, for example via state entropy 
in the pure case.

Acknowledgement: We thank Michael P. Seevinck for pointing out the need for a 
more detailed analysis than previously carried out (in Ref. 8) in order definitively to demonstrate BNSD.

\end{document}